\documentclass[11pt]{article}%
\usepackage{amsfonts}
\usepackage{amsmath}
\usepackage{amssymb}
\usepackage{charter}
\usepackage{graphicx}
\usepackage{cite}
\usepackage[onehalfspacing]{setspace}%
\providecommand{\U}[1]{\protect \rule{.1in}{.1in}}
\RequirePackage{CJK}
\AtBeginDocument{\begin{CJK*}{GBK}{song}\CJKtilde}
\AtEndDocument{\end{CJK*}}
\begin{document}

\title{Two-photon spin states and entanglement states}
\author{Xiao-Jing Liu$^{a, b}$, Xiang-Yao Wu$^{b}$\thanks{E-mail: wuxy2066@163.com},
Jing-Bin Lu$^{a}$, Si-Qi Zhang$^{b}$, Hong-Li$^{b}$\\Ji Ma$^{b}$, Hong-Chun Yuan$^{c}$, Heng-Mei Li$^{c}$, Hai-Xin Gao$^{d}$ and
Jing-Wu Li$^{e}$ \\{\small a) Institute of Physics, Jilin University, Changchun 130012 China}\\{\small b) Institute of Physics, Jilin Normal University, Siping 136000 China}\\{\small c) College of Optoelectronic Engineering, Changzhou Institute of
Technology, Changzhou 213002, China}\\{\small d) Institute of Physics, Northeast Normal University, Changchun 130024
China}\\{\small e) Institute of Physics, Xuzhou Normal University, Xuzhou 221000
China}}
\maketitle

\begin{abstract}
In this paper, we have given the spin states of two-photon, which are
expressed by the quadratic combination of two single photon spin states, they
are a kind of quantum expression. Otherwise, we give all entanglement states
of two-photon, which are different from the xanzsd classical polarization
vector expression of two-photon entanglement state. \newline PACS: 42.50.-p,
42.50.Dv, 03.65.Ud \newline Keywords: two-photon; spin states; entanglement states

\end{abstract}

\section{Introduction}

Quantum entanglement, and its inherent non local properties, are among the
most fascinating and challenging features of the quantum world. In addition,
entanglement plays a central role in quantum information\cite{s1,s2,s3,s4,s5}.
Since its first description in the decade of 1930 \cite{s6}, and in spite of
the decisive contribution of Bell \cite{s7} and the subsequent experimental
studies \cite{s8}, entanglement stays even now as a rather mysterious and
puzzling property of bipartite quantum objects. Entanglement is not only a
fundamental concept in Quantum Mechanics with profound implications, but also
a basic ingredient of many recent technological applications that has been put
forward in quantum communications and quantum computing \cite{s9,s10}.
Entanglement is a very special type of correlation between particles that can
exist in spite of how distant they are. Nevertheless, the term entanglement is
sometimes also used to refer to certain correlations existing between
different degrees of freedom of a single particle \cite{s11}. By and large,
the most common method to generate photonic entanglement, that is entanglement
between photons, is the process of spontaneous parametric down-conversion
(SPDC) \cite{s12}. In SPDC, two lower-frequency photons are generated when an
intense higher-frequency pump beam interacts with the atoms of a
non-centrosymetric nonlinear crystal. Entanglement can reside in any of the
degrees of freedoms that characterize light: angular momentum (polarization
and orbital angular momentum), momentum and frequency, or in several of them,
what is known as hyper-entanglement. Undoubtedly, polarization is the most
widely used resource to generate entanglement between photons thanks to the
existence of many optical elements to control the polarization of light and to
the easiness of its manipulation when compared to other characteristics of a
light beam, e.g., its spatial shape or bandwidth.

Entangled states of photons are the basic resource in the successful
implementation of quantum information processing applications, namely optical
quantum computing \cite{s13,s14}, and quantum cryptography, or quantum key
distribution \cite{s15,s16}. Also, in the experimental study of fundamental
problems in Quantum Mechanics, as loophole free tests of the violation of the
Bell's inequalities \cite{s17}, the delayed-choice quantum eraser \cite{s18},
quantum teleportation and entanglement swapping\cite{s19}, generation of
states with a large number of particles and generalized types of entanglement
\cite{s20}, etc. In this paper, we have given the spin states of two-photon,
which are expressed by the quadratic combination of two single photon spin
states, they are a kind of quantum expression. Otherwise, we give all
entanglement states of two-photon, which are different from the xanzsd
classical polarization vector expression of two-photon entanglement state.

\section{The spin state of two-photon}

The following will give some comments on the real meaning of the state of
photon. In quantum electrodynamics, the spin vector operator of the photon is
\cite{s21,s22} (in natural unit system $\hbar=c=1$)
\begin{equation}
s_{x}=\left(
\begin{array}
[c]{ccc}%
0 & 0 & 0\\
0 & 0 & -i\\
0 & i & 0
\end{array}
\right)  ,s_{y}=\left(
\begin{array}
[c]{ccc}%
0 & 0 & i\\
0 & 0 & 0\\
-i & 0 & 0
\end{array}
\right)  ,s_{z}=\left(
\begin{array}
[c]{ccc}%
0 & -i & 0\\
i & 0 & 0\\
0 & 0 & 0
\end{array}
\right)  ,\label{1}%
\end{equation}%
\begin{equation}
\vec{S}\hspace{0.01in}^{2}=s_{x}^{2}+s_{y}^{2}+s_{z}^{2}=2\left(
\begin{array}
[c]{ccc}%
1 & 0 & 0\\
0 & 1 & 0\\
0 & 0 & 1
\end{array}
\right)  ,\label{2}%
\end{equation}
the spin vector $\chi_{\mu}$ of $S^{2}$ and $s_{z}$ satisfy the following
equations
\begin{equation}
\vec{S}\hspace{0.01in}^{2}\chi_{\mu}=2\chi_{\mu},\label{3}%
\end{equation}%
\begin{equation}
s_{z}\chi_{\mu}=\mu \chi_{\mu},\label{4}%
\end{equation}
these are
\begin{equation}
\chi_{0}=\left(
\begin{array}
[c]{c}%
0\\
0\\
1
\end{array}
\right)  ,\chi_{1}=-\frac{1}{\sqrt{2}}\left(
\begin{array}
[c]{c}%
1\\
i\\
0
\end{array}
\right)  ,\chi_{-1}=\frac{1}{\sqrt{2}}\left(
\begin{array}
[c]{c}%
1\\
-i\\
0
\end{array}
\right)  ,\label{5}%
\end{equation}
the total spin $\vec{S}$ is the sum of two photons spin $s_{1}$ and $s_{2}$
\begin{equation}
\vec{S}=\vec{s}_{1}+\vec{s}_{2},\label{6}%
\end{equation}
and $\vec{S}\hspace{0.01in}^{2}$ eigenvalue is
\begin{equation}
\vec{S}\hspace{0.01in}^{2}=S(S+1),\label{7}%
\end{equation}
and quantum number $S$ are
\begin{equation}
S=0,1,2.\label{8}%
\end{equation}
For the given $S$, $s_{z}$ eigenvalues are: $\mu=-S,-S+1,\cdot \cdot \cdot S$,
i.e., there are $2S+1$ eigenvalues, and $2S+1$ eigenfunctions of spin. For
$S=0$, there is one eigenfunction, for $S=1$, there are three eigenfunctions,
and for $S=2$, there are five eigenfunctions, i.e., there are nine spin
wave-functions $\chi_{S\mu}$ for two-photon, which are expressed by the
quadratic combination of two single photon spin wave functions $\chi_{\mu1}$
and $\chi_{\mu2}$. We try to write the nine spin wave functions of two-photon,
they are\newline(1) $S=2$ spin wave functions
\begin{equation}
\chi_{S}^{(1)}=\chi_{22}(s_{1z},s_{2z})=\chi_{1}(s_{1z})\chi_{1}%
(s_{2z}),\hspace{0.3in}(S=2,\mu=2)\label{9}%
\end{equation}%
\begin{equation}
\chi_{S}^{(2)}=\chi_{21}(s_{1z},s_{2z})=\frac{1}{\sqrt{2}}[\chi_{0}%
(s_{1z})\chi_{1}(s_{2z})+\chi_{0}(s_{2z})\chi_{1}(s_{1z})],\hspace
{0.3in}(S=2,\mu=1)\label{10}%
\end{equation}%
\begin{equation}
\chi_{S}^{(3)}=\chi_{20}(s_{1z},s_{2z})=\frac{1}{\sqrt{2}}[\chi_{1}%
(s_{1z})\chi_{-1}(s_{2z})+\chi_{1}(s_{2z})\chi_{-1}(s_{1z})],\hspace
{0.3in}(S=2,\mu=0)\label{11}%
\end{equation}%
\begin{equation}
\chi_{S}^{(4)}=\chi_{2-1}(s_{1z},s_{2z})=\frac{1}{\sqrt{2}}[\chi_{0}%
(s_{1z})\chi_{-1}(s_{2z})+\chi_{0}(s_{2z})\chi_{-1}(s_{1z})],\hspace
{0.3in}(S=2,\mu=-1)\label{12}%
\end{equation}%
\begin{equation}
\chi_{S}^{(5)}=\chi_{2-2}(s_{1z},s_{2z})=\chi_{-1}(s_{1z})\chi_{-1}%
(s_{2z}),\hspace{0.3in}(S=2,\mu=-2)\label{13}%
\end{equation}
(2) $S=0$ spin wave function
\begin{equation}
\chi_{S}^{(6)}=\chi_{00}(s_{1z},s_{2z})=\chi_{0}(s_{1z})\chi_{0}%
(s_{2z}),\hspace{0.3in}(S=0,\mu=0)\label{14}%
\end{equation}
(3) $S=1$ spin wave functions
\begin{equation}
\chi_{A}^{(1)}=\chi_{11}(s_{1z},s_{2z})=\frac{1}{\sqrt{2}}[\chi_{0}%
(s_{1z})\chi_{1}(s_{2z})-\chi_{0}(s_{2z})\chi_{1}(s_{1z})],\hspace
{0.3in}(S=1,\mu=1)\label{15}%
\end{equation}%
\begin{equation}
\chi_{A}^{(2)}=\chi_{10}(s_{1z},s_{2z})=\frac{1}{\sqrt{2}}[\chi_{1}%
(s_{1z})\chi_{-1}(s_{2z})-\chi_{1}(s_{2z})\chi_{-1}(s_{1z})],\hspace
{0.3in}(S=1,\mu=0)\label{16}%
\end{equation}%
\begin{equation}
\chi_{A}^{(3)}=\chi_{1-1}(s_{1z},s_{2z})=\frac{1}{\sqrt{2}}[\chi_{0}%
(s_{1z})\chi_{-1}(s_{2z})-\chi_{0}(s_{2z})\chi_{-1}(s_{1z})],\hspace
{0.3in}(S=1,\mu=-1)\label{17}%
\end{equation}
In the following, we shall check the above spin wave functions, the two-photon
total spin square $\vec{S}\hspace{0.01in}^{2}$ and its $z$ component $s_{z}$
are
\begin{align}
\vec{S}\hspace{0.01in}^{2}=(\vec{s_{1}}+\vec{s_{2}})\hspace{0.0005in}^{2}  &
=\vec{s}_{1}\hspace{0.0005in}^{2}+\vec{s}_{2}\hspace{0.0005in}^{2}%
+2(s_{1x}s_{2x}+s_{1y}s_{2y}+s_{1z}s_{2z})\nonumber \\
& =4+2(s_{1x}s_{2x}+s_{1y}s_{2y}+s_{1z}s_{2z}),\label{18}%
\end{align}%
\begin{equation}
s_{z}=s_{1z}+s_{2z},\label{19}%
\end{equation}
by spin wave functions (\ref{5}), we have
\begin{equation}
s_{x}\chi_{0}=\left(
\begin{array}
[c]{ccc}%
0 & 0 & 0\\
0 & 0 & -i\\
0 & i & 0
\end{array}
\right)  \left(
\begin{array}
[c]{c}%
0\\
0\\
1
\end{array}
\right)  =\frac{1}{\sqrt{2}}(\chi_{1}+\chi_{-1}),\label{20}%
\end{equation}%
\begin{equation}
s_{x}\chi_{1}=-\frac{1}{\sqrt{2}}\left(
\begin{array}
[c]{ccc}%
0 & 0 & 0\\
0 & 0 & -i\\
0 & i & 0
\end{array}
\right)  \left(
\begin{array}
[c]{c}%
1\\
i\\
0
\end{array}
\right)  =\frac{1}{\sqrt{2}}\chi_{0},\label{21}%
\end{equation}%
\begin{equation}
s_{x}\chi_{-1}=\frac{1}{\sqrt{2}}\left(
\begin{array}
[c]{ccc}%
0 & 0 & 0\\
0 & 0 & -i\\
0 & i & 0
\end{array}
\right)  \left(
\begin{array}
[c]{c}%
1\\
-i\\
0
\end{array}
\right)  =\frac{1}{\sqrt{2}}\chi_{0},\label{22}%
\end{equation}%
\begin{equation}
s_{y}\chi_{0}=-\frac{i}{\sqrt{2}}(\chi_{1}-\chi_{-1}),\label{23}%
\end{equation}%
\begin{equation}
s_{y}\chi_{1}=\frac{i}{\sqrt{2}}\chi_{0},\label{24}%
\end{equation}%
\begin{equation}
s_{y}\chi_{-1}=-\frac{i}{\sqrt{2}}\chi_{0},\label{25}%
\end{equation}%
\begin{equation}
s_{z}\chi_{0}=0\cdot \chi_{0},\label{26}%
\end{equation}%
\begin{equation}
s_{z}\chi_{1}=1\cdot \chi_{1},\label{27}%
\end{equation}%
\begin{equation}
s_{z}\chi_{-1}=-1\cdot \chi_{-1},\label{28}%
\end{equation}
with equations (\ref{20})-(\ref{28}), we can check spin wave
functions.\newline(1) checking $S=2$ spin wave functions

(a) checking spin wave function $\chi_{S}^{(1)}$:
\begin{align}
s_{z}\chi_{S}^{(1)}  & =(s_{1z}+s_{2z})\chi_{1}(s_{1z})\chi_{1}(s_{2z}%
)\nonumber \\
& =(s_{1z}\chi_{1}(s_{1z}))\chi_{1}(s_{2z})+\chi_{1}(s_{1z})(s_{2z}\chi
_{1}(s_{2z}))\nonumber \\
& =2\chi_{1}(s_{1z})\chi_{1}(s_{2z}),\label{29}%
\end{align}
i.e.,
\begin{equation}
\mu=2,\label{30}%
\end{equation}%
\begin{align}
\vec{S}^{2}\chi_{S}^{(1)}  & =[4+2(s_{1x}s_{2x}+s_{1y}s_{2y}+s_{1z}%
s_{2z})]\chi_{1}(s_{1z})\chi_{1}(s_{2z})\nonumber \\
& =4\chi_{1}(s_{1z})\chi_{1}(s_{2z})+2[\frac{1}{\sqrt{2}}\chi_{0}(s_{1z}%
)\frac{1}{\sqrt{2}}\chi_{0}(s_{2z})+\frac{i}{\sqrt{2}}\chi_{0}(s_{1z})\frac
{i}{\sqrt{2}}\chi_{0}(s_{2z})+\chi_{1}(s_{1z})\chi_{1}(s_{2z})]\nonumber \\
& =6\chi_{1}(s_{1z})\chi_{1}(s_{2z}),\label{31}%
\end{align}
i.e.,
\begin{equation}
S=2,\label{32}%
\end{equation}
we have checked $\chi_{S}^{(1)}$ is the spin wave function of two-photon
corresponding to $S=2$ and $\mu=2$.

(b) checking spin wave function $\chi_{S}^{(2)}$:
\begin{align}
s_{z}\chi_{S}^{(2)}  & =(s_{1z}+s_{2z})\frac{1}{\sqrt{2}}[\chi_{0}(s_{1z}%
)\chi_{1}(s_{2z})+\chi_{0}(s_{2z})\chi_{1}(s_{1z})]\nonumber \\
& =\frac{1}{\sqrt{2}}[\chi_{0}(s_{2z})\chi_{1}(s_{1z})+\chi_{0}(s_{1z}%
)\chi_{1}(s_{2z})],\label{33}%
\end{align}
i.e.,
\begin{equation}
\mu=1,\label{34}%
\end{equation}%
\begin{align}
\vec{S}^{2}\chi_{S}^{(2)}  & =[4+2(s_{1x}s_{2x}+s_{1y}s_{2y}+s_{1z}%
s_{2z})]\frac{1}{\sqrt{2}}[\chi_{0}(s_{1z})\chi_{1}(s_{2z})+\chi_{0}%
(s_{2z})\chi_{1}(s_{1z})]\nonumber \\
& =2\sqrt{2}[\chi_{0}(s_{1z})\chi_{1}(s_{2z})+\chi_{0}(s_{2z})\chi_{1}%
(s_{1z})]\nonumber \\
& +\sqrt{2}[\frac{1}{2}(\chi_{1}(s_{1z})\chi_{0}(s_{2z})+\chi_{-1}(s_{1z}%
)\chi_{0}(s_{2z}))+\frac{1}{2}(\chi_{1}(s_{2z})\chi_{0}(s_{1z})+\chi
_{-1}(s_{2z})\chi_{0}(s_{1z}))]\nonumber \\
& +\sqrt{2}[\frac{1}{2}(\chi_{1}(s_{1z})\chi_{0}(s_{2z})-\chi_{-1}(s_{1z}%
)\chi_{0}(s_{2z}))+\frac{1}{2}(\chi_{1}(s_{2z})\chi_{0}(s_{1z})-\chi
_{-1}(s_{2z})\chi_{0}(s_{1z}))]\nonumber \\
& =6\cdot \frac{1}{\sqrt{2}}(\chi_{0}(s_{1z})\chi_{1}(s_{2z})+\chi_{0}%
(s_{2z})\chi_{1}(s_{1z})),\label{35}%
\end{align}
i.e.,
\begin{equation}
S=2,\label{36}%
\end{equation}
we have checked $\chi_{S}^{(2)}$ is the spin wave function of two-photon
corresponding to $S=2$ and $\mu=1$.

(c) checking spin wave function $\chi_{S}^{(3)}$:
\begin{align}
s_{z}\chi_{S}^{(3)}  & =(s_{1z}+s_{2z})\frac{1}{\sqrt{2}}[\chi_{1}(s_{1z}%
)\chi_{-1}(s_{2z})+\chi_{1}(s_{2z})\chi_{-1}(s_{1z})]\nonumber \\
& =\frac{1}{\sqrt{2}}[(s_{1z}\chi_{1}(s_{1z}))\chi_{-1}(s_{2z})+\chi
_{1}(s_{2z})(s_{1z}\chi_{-1}(s_{1z})]\nonumber \\
& +\frac{1}{\sqrt{2}}[\chi_{1}(s_{1z})(s_{2z}\chi_{-1}(s_{2z}))+(s_{2z}%
\chi_{1}(s_{2z}))\chi_{-1}(s_{1z})]\nonumber \\
& =0,\label{37}%
\end{align}
i.e.,
\begin{equation}
\mu=0,\label{38}%
\end{equation}%
\begin{align}
\vec{S}^{2}\chi_{S}^{(3)}  & =[4+2(s_{1x}s_{2x}+s_{1y}s_{2y}+s_{1z}%
s_{2z})]\frac{1}{\sqrt{2}}[\chi_{1}(s_{1z})\chi_{-1}(s_{2z})+\chi_{1}%
(s_{2z})\chi_{-1}(s_{1z})]\nonumber \\
& =2\sqrt{2}[\chi_{1}(s_{1z})\chi_{-1}(s_{2z})+\chi_{1}(s_{2z})\chi
_{-1}(s_{1z})]\nonumber \\
& +\sqrt{2}[\frac{1}{\sqrt{2}}\chi_{0}(s_{1z})\frac{1}{\sqrt{2}}\chi
_{0}(s_{1z}))+\frac{1}{\sqrt{2}}\chi_{0}(s_{2z})\frac{1}{\sqrt{2}}\chi
_{0}(s_{1z}))]\nonumber \\
& +\sqrt{2}[\frac{i}{\sqrt{2}}\chi_{0}(s_{1z})(-\frac{1}{\sqrt{2}})\chi
_{0}(s_{2z}))+\frac{i}{\sqrt{2}}\chi_{0}(s_{2z})(-\frac{1}{\sqrt{2}})\chi
_{0}(s_{1z}))]\nonumber \\
& +\sqrt{2}[-\chi_{1}(s_{1z})\chi_{-1}(s_{2z}))-\chi_{1}(s_{2z})\chi
_{-1}(s_{1z}))]\nonumber \\
& =2\cdot \frac{1}{\sqrt{2}}[\chi_{1}(s_{1z})\chi_{-1}(s_{2z})+\chi_{1}%
(s_{2z})\chi_{-1}(s_{1z})]\nonumber \\
& +2\cdot \frac{1}{\sqrt{2}}[\chi_{0}(s_{1z})\chi_{0}(s_{2z})+\chi_{0}%
(s_{2z})\chi_{0}(s_{1z})],\label{39}%
\end{align}
we find that $\chi_{S}^{(3)}$ is not the common eigenstate of $\{ \vec
{S}\hspace{0.005in}^{2},s_{z}\}$, corresponding to $S=2$ and $\mu=0$. Their
common eigenstate $\chi_{S}^{(3)}$ can be obtained by the linear superposition
of $\chi_{20}$ and $\chi_{00}$, the $\chi_{S}^{(3)}$ can be written as
\begin{equation}
\chi_{S}^{(3)}=a\chi_{0}(s_{1z})\chi_{0}(s_{2z})+b[\chi_{1}(s_{1z})\chi
_{-1}(s_{2z})+\chi_{1}(s_{2z})\chi_{-1}(s_{1z})],\label{40}%
\end{equation}
by the eigenequations
\begin{equation}
\left \{
\begin{array}
[c]{ll}%
\vec{S}^{2}\chi_{S}^{(3)}=6\chi_{S}^{(3)} & \\
s_{z}\chi_{S}^{(3)}=0 &
\end{array}
\right.  ,\label{41}%
\end{equation}
we can calculate the superposition coefficients $a$ and $b$
\begin{align}
s_{z}\chi_{S}^{(3)}  & =a(s_{1z}\chi_{0}(s_{1z}))\chi_{0}(s_{2z}%
)+b[(s_{1z}\chi_{1}(s_{1z}))\chi_{-1}(s_{2z})+(s_{1z}\chi_{-1}(s_{1z}%
))\chi_{1}(s_{2z})]\nonumber \\
& +a(s_{2z}\chi_{0}(s_{2z}))\chi_{0}(s_{1z})+b[(s_{2z}\chi_{-1}(s_{2z}%
))\chi_{1}(s_{1z})+(s_{2z}\chi_{1}(s_{2z}))\chi_{-1}(s_{1z})]\nonumber \\
& =b[\chi_{1}(s_{1z})\chi_{-1}(s_{2z})-\chi_{1}(s_{2z})\chi_{-1}%
(s_{1z})]+b[-\chi_{1}(s_{1z})\chi_{-1}(s_{2z})+\chi_{1}(s_{2z})\chi
_{-1}(s_{1z})]\nonumber \\
& =0,\label{42}%
\end{align}
i.e.,
\begin{equation}
\mu \equiv0,\label{43}%
\end{equation}%
\begin{align}
\vec{S^{2}}\chi_{S}^{(3)}  & =4[a\chi_{0}(s_{1z})\chi_{0}(s_{2z})+b(\chi
_{1}(s_{1z})\chi_{-1}(s_{2z})+\chi_{1}(s_{2z})\chi_{-1}(s_{1z}))]\nonumber \\
& +2[a(s_{1x}\chi_{0}(s_{1z}))(s_{2x}\chi_{0}(s_{2z}))+b((s_{1x}\chi
_{1}(s_{1z}))(s_{2x}\chi_{-1}(s_{2z}))+(s_{1x}\chi_{-1}(s_{1z}))(s_{2x}%
\chi_{1}(s_{2z})))]\nonumber \\
& +2[a(s_{1y}\chi_{0}(s_{1z}))(s_{2y}\chi_{0}(s_{2z}))+b((s_{1y}\chi
_{1}(s_{1z}))(s_{2y}\chi_{-1}(s_{2z}))+(s_{1y}\chi_{-1}(s_{1z}))(s_{2y}%
\chi_{1}(s_{2z})))]\nonumber \\
& +2[a(s_{1z}\chi_{0}(s_{1z}))(s_{2z}\chi_{0}(s_{2z}))+b((s_{1z}\chi
_{1}(s_{1z}))(s_{2z}\chi_{-1}(s_{2z}))+(s_{1z}\chi_{-1}(s_{1z}))(s_{2z}%
\chi_{1}(s_{2z})))]\nonumber \\
& +2[b(-\chi_{1}(s_{1z}))\chi_{-1}(s_{2z})-(\chi_{-1}(s_{1z}))\chi_{1}%
(s_{2z})]\nonumber \\
& =4[a\chi_{0}(s_{1z})\chi_{0}(s_{2z})+b(\chi_{1}(s_{1z})\chi_{-1}%
(s_{2z})+\chi_{1}(s_{2z})\chi_{-1}(s_{1z}))]+2[a(\chi_{1}(s_{1z})\chi
_{-1}(s_{2z})\nonumber \\
& +\chi_{1}(s_{2z})\chi_{-1}(s_{1z}))+2b\chi_{0}(s_{1z})\chi_{0}%
(s_{2z})-b(\chi_{1}(s_{1z})\chi_{-1}(s_{2z})+\chi_{1}(s_{2z})\chi_{-1}%
(s_{1z}))]\nonumber \\
& =4(a+b)\chi_{0}(s_{1z})\chi_{0}(s_{2z})+2(a+b)(\chi_{1}(s_{1z})\chi
_{-1}(s_{2z})+\chi_{1}(s_{2z})\chi_{-1}(s_{1z}))\nonumber \\
& =6[a\chi_{0}(s_{1z})\chi_{0}(s_{2z})+b(\chi_{1}(s_{1z})\chi_{-1}%
(s_{2z})+\chi_{1}(s_{2z})\chi_{-1}(s_{1z}))],\label{44}%
\end{align}
comparing the coefficients of the same terms, we obtain
\begin{equation}
\left \{
\begin{array}
[c]{ll}%
4(a+b)=6a & \\
2(a+b)=6b &
\end{array}
\right.  ,\label{45}%
\end{equation}
i.e.,
\begin{equation}
a=2b,\label{46}%
\end{equation}
then
\begin{equation}
\chi_{S}^{(3)}=2b\chi_{0}(s_{1z})\chi_{0}(s_{2z})+b[\chi_{1}(s_{1z})\chi
_{-1}(s_{2z})+\chi_{1}(s_{2z})\chi_{-1}(s_{1z})],\label{47}%
\end{equation}
by condition of normalization
\begin{equation}
(\chi_{S}^{(3)})^{+}\chi_{S}^{(3)}=1,\label{48}%
\end{equation}
we have
\begin{equation}
b=\frac{1}{\sqrt{6}},\label{49}%
\end{equation}
we get the eigenstate $\chi_{S}^{(3)}$, it is
\begin{equation}
\chi_{S}^{(3)}=\frac{2}{\sqrt{6}}\chi_{0}(s_{1z})\chi_{0}(s_{2z})+\frac
{1}{\sqrt{6}}[\chi_{1}(s_{1z})\chi_{-1}(s_{2z})+\chi_{1}(s_{2z})\chi
_{-1}(s_{1z})],\label{50}%
\end{equation}

(d) checking spin wave function $\chi_{S}^{(4)}$:
\begin{align}
s_{z}\chi_{S}^{(4)}  & =\frac{1}{\sqrt{2}}[(s_{1z}\chi_{0}(s_{1z}))\chi
_{-1}(s_{2z})+\chi_{0}(s_{2z})(s_{1z}\chi_{-1}(s_{1z}))]\nonumber \\
& +\frac{1}{\sqrt{2}}[\chi_{0}(s_{1z})(s_{2z}\chi_{-1}(s_{2z}))+(s_{2z}%
\chi_{0}(s_{2z}))\chi_{-1}(s_{1z})]\nonumber \\
& =-\frac{1}{\sqrt{2}}[\chi_{0}(s_{1z})\chi_{-1}(s_{2z})+\chi_{0}(s_{2z}%
)\chi_{-1}(s_{1z})],\label{51}%
\end{align}
i.e.,
\begin{equation}
\mu=-1,\label{52}%
\end{equation}%
\begin{align}
\vec{S^{2}}\chi_{S}^{(4)}  & =4\chi_{S}^{(4)}+2\frac{1}{\sqrt{2}}[(s_{1x}%
\chi_{0}(s_{1z}))(s_{2x}\chi_{-1}(s_{2z}))+(s_{2x}\chi_{0}(s_{2z}))(s_{1x}%
\chi_{-1}(s_{1z}))]\nonumber \\
& +2\frac{1}{\sqrt{2}}[(s_{1y}\chi_{0}(s_{1z}))(s_{2y}\chi_{-1}(s_{2z}%
))+(s_{2y}\chi_{0}(s_{2z}))(s_{1y}\chi_{-1}(s_{1z}))]\nonumber \\
& +2\frac{1}{\sqrt{2}}[(s_{1z}\chi_{0}(s_{1z}))(s_{2z}\chi_{-1}(s_{2z}%
))+(s_{2z}\chi_{0}(s_{2z}))(s_{1z}\chi_{-1}(s_{1z}))]\nonumber \\
& =4\chi_{S}^{(4)}+2\frac{1}{\sqrt{2}}[\chi_{-1}(s_{1z})\chi_{0}(s_{2z}%
)+\chi_{-1}(s_{2z})\chi_{0}(s_{1z})]\nonumber \\
& =6\chi_{S}^{(4)},\label{53}%
\end{align}
i.e.,
\begin{equation}
S=2,\label{54}%
\end{equation}
we have checked $\chi_{S}^{(4)}$ is the spin wave function of two-photon
corresponding to $S=2$ and $\mu=-1$.

(e) checking spin wave function $\chi_{S}^{(5)}$:
\begin{align}
s_{z}\chi_{S}^{(5)}  & =(s_{1z}\chi_{-1}(s_{1z}))\chi_{-1}(s_{2z})+\chi
_{-1}(s_{1z})(s_{2z}\chi_{-1}(s_{2z}))\nonumber \\
& =-\chi_{1}(s_{1z})\chi_{-1}(s_{2z})-\chi_{-1}(s_{1z})\chi_{-1}%
(s_{2z})\nonumber \\
& =-2[\chi_{1}(s_{1z})\chi_{-1}(s_{2z})],\label{55}%
\end{align}
i.e.,
\begin{equation}
\mu=-2,\label{56}%
\end{equation}%
\begin{align}
\vec{S}^{2}\chi_{S}^{(5)}  & =4\chi_{-1}(s_{1z})\chi_{-1}(s_{2z}%
)+2[(s_{1x}\chi_{-1}(s_{1z}))(s_{2x}\chi_{-1}(s_{2z}))\nonumber \\
& +(s_{1y}\chi_{-1}(s_{1z}))(s_{2y}\chi_{-1}(s_{2z}))+(s_{1z}\chi_{-1}%
(s_{1z}))(s_{2z}\chi_{-1}(s_{2z}))]\nonumber \\
& =4\chi_{-1}(s_{1z})\chi_{-1}(s_{2z})+2[\frac{1}{\sqrt{2}}\chi_{0}%
(s_{1z})\frac{1}{\sqrt{2}}\chi_{0}(s_{2z})+\frac{-i}{\sqrt{2}}\chi_{0}%
(s_{1z})\frac{-i}{\sqrt{2}}\chi_{0}(s_{2z})\nonumber \\
& +\chi_{-1}(s_{1z})\chi_{-1}(s_{2z})]\nonumber \\
& =6\chi_{-1}(s_{1z})\chi_{-1}(s_{2z}),\label{57}%
\end{align}
i.e.,
\begin{equation}
S=2,\label{58}%
\end{equation}
we have checked $\chi_{S}^{(5)}$ is the spin wave function of two-photon
corresponding to $S=2$ and $\mu=-2$.\newline(2) checking $S=0$ spin wave
functions $\chi_{S}^{(6)}$:
\begin{equation}
s_{z}\chi_{s}^{(6)}=(s_{1z}\chi_{0}(s_{1z}))\chi_{0}(s_{2z})+\chi_{0}%
(s_{1z})(s_{2z}\chi_{0}(s_{2z}))=0,\label{59}%
\end{equation}
i.e.,
\begin{equation}
\mu=0,\label{60}%
\end{equation}%
\begin{align}
\vec{S}^{2}\chi_{S}^{(6)}  & =4\chi_{0}(s_{1z})\chi_{0}(s_{2z})+2[(s_{1x}%
\chi_{0}(s_{1z}))(s_{2x}\chi_{0}(s_{2z}))\nonumber \\
& +(s_{1y}\chi_{0}(s_{1z}))(s_{2y}\chi_{0}(s_{2z}))+(s_{1z}\chi_{0}%
(s_{1z}))(s_{2z}\chi_{0}(s_{2z}))]\nonumber \\
& =4\chi_{0}(s_{1z})\chi_{0}(s_{2z})+2[\frac{1}{\sqrt{2}}(\chi_{1}%
(s_{1z})+\chi_{-1}(s_{1z}))\frac{1}{\sqrt{2}}(\chi_{1}(s_{2z})+\chi
_{-1}(s_{2z}))\nonumber \\
& +\frac{-i}{\sqrt{2}}(\chi_{1}(s_{1z})-\chi_{-1}(s_{1z}))\frac{-i}{\sqrt{2}%
}(\chi_{1}(s_{2z})-\chi_{-1}(s_{2z}))]\nonumber \\
& =4\chi_{0}(s_{1z})\chi_{0}(s_{2z})+2[\chi_{1}(s_{1z})\chi_{-1}(s_{2z}%
)+\chi_{1}(s_{2z})\chi_{-1}(s_{1z})]\neq0,\label{61}%
\end{align}
we find $\chi_{S}^{(6)}$ is not the common eigenstate of $\{ \vec{S}%
\hspace{0.005in}^{2},s_{z}\}$, corresponding to $S=0$ and $\mu=0$. Their
common eigenstate $\chi_{S}^{(6)}$ can be written as equation (\ref{40}), we
have
\begin{equation}
s_{z}\chi_{S}^{(6)}=0,\label{62}%
\end{equation}
i.e.,
\begin{equation}
\mu \equiv0,\label{63}%
\end{equation}
and
\begin{align}
\vec{S}^{2}\chi_{S}^{(6)}  & =4[a\chi_{0}(s_{1z})\chi_{0}(s_{2z})+b(\chi
_{1}(s_{1z})\chi_{-1}(s_{2z})+\chi_{1}(s_{2z})\chi_{-1}(s_{1z}))]\nonumber \\
& +4b\chi_{0}(s_{1z})\chi_{0}(s_{2z})+2(a-b)(\chi_{1}(s_{1z})\chi_{-1}%
(s_{2z})+\chi_{1}(s_{2z})\chi_{-1}(s_{1z}))\nonumber \\
& =4(a+b)\chi_{0}(s_{1z})\chi_{0}(s_{2z})+2(a+b)(\chi_{1}(s_{1z})\chi
_{-1}(s_{2z})+\chi_{1}(s_{2z})\chi_{-1}(s_{1z})))\nonumber \\
& =0,\label{64}%
\end{align}
we have
\begin{equation}
\left \{
\begin{array}
[c]{ll}%
4(a+b)=0 & \\
2(a+b)=0 &
\end{array}
\right.  ,\label{65}%
\end{equation}
i.e.,
\begin{equation}
a=-b,\label{66}%
\end{equation}
then
\begin{equation}
\chi_{S}^{(6)}=a\chi_{0}(s_{1z})\chi_{0}(s_{2z})-a[\chi_{1}(s_{1z})\chi
_{-1}(s_{2z})+\chi_{1}(s_{2z})\chi_{-1}(s_{1z})],\label{67}%
\end{equation}
by condition of normalization
\begin{equation}
(\chi_{S}^{(6)})^{+}\chi_{S}^{(6)}=1,\label{68}%
\end{equation}%
\begin{equation}
a=\frac{1}{\sqrt{3}},\label{69}%
\end{equation}
we get the eigenstate $\chi_{S}^{(6)}$, it is
\begin{equation}
\chi_{S}^{(6)}=\frac{1}{\sqrt{3}}\chi_{0}(s_{1z})\chi_{0}(s_{2z})-\frac
{1}{\sqrt{3}}[\chi_{1}(s_{1z})\chi_{-1}(s_{2z})+\chi_{1}(s_{2z})\chi
_{-1}(s_{1z})].\label{70}%
\end{equation}
(3) $S=1$ spin wave functions

(a) checking spin wave function $\chi_{A}^{(1)}$:
\begin{align}
s_{z}X_{A}^{(1)}  & =(s_{1z}+s_{2z})\frac{1}{\sqrt{2}}[\chi_{0}(s_{1z}%
)\chi_{1}(s_{2z})-\chi_{0}(s_{2z})\chi_{1}(s_{1z})]\nonumber \\
& =\frac{1}{\sqrt{2}}[(s_{1z}\chi_{0}(s_{1z}))\chi_{1}(s_{2z})-\chi_{0}%
(s_{2z})(s_{1z}\chi_{1}(s_{1z}))]\nonumber \\
& +\frac{1}{\sqrt{2}}[\chi_{0}(s_{1z})(s_{2z}\chi_{1}(s_{2z}))-(s_{2z}\chi
_{0}(s_{2z}))\chi_{1}(s_{1z})]\nonumber \\
& =\frac{1}{\sqrt{2}}[\chi_{0}(s_{1z})\chi_{1}(s_{2z})-\chi_{0}(s_{2z}%
)\chi_{1}(s_{1z})],\label{71}%
\end{align}
i.e.,
\begin{equation}
\mu=1,\label{72}%
\end{equation}%
\begin{align}
\vec{S}^{2}\chi_{A}^{(1)}  & =4\chi_{A}^{(1)}+\sqrt{2}[(s_{1x}\chi_{0}%
(s_{1z}))(s_{2x}\chi_{1}(s_{2z}))-(s_{2x}\chi_{0}(s_{2z}))(s_{1x}\chi
_{1}(s_{1z}))]\nonumber \\
& +\sqrt{2}[(s_{1y}\chi_{0}(s_{1z}))(s_{2y}\chi_{1}(s_{2z}))-(s_{2y}\chi
_{0}(s_{2z}))(s_{1y}\chi_{1}(s_{1z}))]\nonumber \\
& +\sqrt{2}[(s_{1z}\chi_{0}(s_{1z}))(s_{2z}\chi_{1}(s_{2z}))-(s_{2z}\chi
_{0}(s_{2z}))(s_{1z}\chi_{1}(s_{1z}))]\nonumber \\
& =4\chi_{A}^{(1)}+\sqrt{2}[\frac{1}{\sqrt{2}}(\chi_{1}(s_{1z})+\chi
_{-1}(s_{1z}))\frac{1}{\sqrt{2}}\chi_{0}(s_{2z})-\frac{1}{\sqrt{2}}(\chi
_{1}(s_{2z})+\chi_{-1}(s_{2z}))\frac{1}{\sqrt{2}}\chi_{0}(s_{1z})]\nonumber \\
& +\sqrt{2}[-\frac{i}{\sqrt{2}}(\chi_{1}(s_{1z})-\chi_{-1}(s_{1z}))\frac
{i}{\sqrt{2}}\chi_{0}(s_{2z})+\frac{i}{\sqrt{2}}(\chi_{1}(s_{2z})-\chi
_{-1}(s_{2z}))\frac{i}{\sqrt{2}}\chi_{0}(s_{1z})]\nonumber \\
& =4\chi_{A}^{(1)}+\sqrt{2}[\chi_{1}(s_{1z})\chi_{0}(s_{2z})-\chi_{1}%
(s_{2z})\chi_{0}(s_{1z})]\nonumber \\
& =4\chi_{A}^{(1)}-2\chi_{A}^{(1)}=2\chi_{A}^{(1)},\label{73}%
\end{align}
i.e.,
\begin{equation}
S=1,\label{74}%
\end{equation}
we have checked $\chi_{A}^{(1)}$ is the spin wave function of two-photon
corresponding to $S=1$ and $\mu=1$.

(b) checking spin wave function $\chi_{A}^{(2)}$:
\begin{align}
s_{z}X_{A}^{(2)}  & =\frac{1}{\sqrt{2}}[(s_{1z}\chi_{1}(s_{1z}))\chi
_{-1}(s_{2z})-\chi_{1}(s_{2z})(s_{1z}\chi_{-1}(s_{1z}))]\nonumber \\
& +\frac{1}{\sqrt{2}}[\chi_{1}(s_{1z})(s_{2z}\chi_{-1}(s_{2z}))-(s_{2z}%
\chi_{1}(s_{2z}))\chi_{-1}(s_{1z})]\nonumber \\
& =0,\label{75}%
\end{align}
i.e.,
\begin{equation}
\mu=0,\label{76}%
\end{equation}%
\begin{align}
\vec{S}^{2}\chi_{A}^{(2)}  & =4\chi_{A}^{(2)}+\sqrt{2}[(s_{1x}\chi_{1}%
(s_{1z}))(s_{2x}\chi_{-1}(s_{2z}))-(s_{2x}\chi_{1}(s_{2z}))(s_{1x}\chi
_{-1}(s_{1z}))]\nonumber \\
& +\sqrt{2}[(s_{1y}\chi_{1}(s_{1z}))(s_{2y}\chi_{-1}(s_{2z}))-(s_{2y}\chi
_{1}(s_{2z}))(s_{1y}\chi_{-1}(s_{1z}))]\nonumber \\
& +\sqrt{2}[(s_{1z}\chi_{1}(s_{1z}))(s_{2z}\chi_{-1}(s_{2z}))-(s_{2z}\chi
_{1}(s_{2z}))(s_{1z}\chi_{-1}(s_{1z}))]\nonumber \\
& =4\chi_{A}^{(2)}+2\cdot \frac{1}{{}}\sqrt{2}[\frac{1}{\sqrt{2}}\chi
_{0}(s_{1z})\frac{1}{\sqrt{2}}\chi_{0}(s_{2z})-\frac{1}{\sqrt{2}}\chi
_{0}(s_{2z})\frac{1}{\sqrt{2}}\chi_{0}(s_{1z})\nonumber \\
& +\frac{i}{\sqrt{2}}\chi_{0}(s_{1z})(-\frac{i}{\sqrt{2}})\chi_{0}%
(s_{2z})-\frac{i}{\sqrt{2}}\chi_{0}(s_{2z})(-\frac{i}{\sqrt{2}})\chi
_{0}(s_{1z})\nonumber \\
& -\chi_{1}(s_{1z})\chi_{-1}(s_{2z})+\chi_{1}(s_{2z})\chi_{-1}(s_{1z}%
)]\nonumber \\
& =4\chi_{A}^{(2)}-2\chi_{A}^{(2)}=2\chi_{A}^{(2)},\label{77}%
\end{align}
i.e.,
\begin{equation}
S=1,\label{78}%
\end{equation}
we have checked $\chi_{A}^{(2)}$ is the spin wave function of two-photon
corresponding to $S=1$ and $\mu=0$.

(c) checking spin wave function $\chi_{A}^{(3)}$:
\begin{align}
s_{z}X_{A}^{(3)}  & =\frac{1}{\sqrt{2}}[(s_{1z}\chi_{0}(s_{1z}))\chi
_{-1}(s_{2z})-\chi_{0}(s_{2z})(s_{1z}\chi_{-1}(s_{1z}))]\nonumber \\
& +\frac{1}{\sqrt{2}}[\chi_{0}(s_{1z})(s_{2z}\chi_{-1}(s_{2z}))-(s_{2z}%
\chi_{0}(s_{2z}))\chi_{-1}(s_{1z})]\nonumber \\
& =-\frac{1}{\sqrt{2}}[\chi_{0}(s_{1z})\chi_{-1}(s_{2z})-\chi_{0}(s_{2z}%
)\chi_{-1}(s_{1z})],\label{79}%
\end{align}
i.e.,
\begin{equation}
\mu=-1,\label{80}%
\end{equation}%
\begin{align}
\vec{S}^{2}\chi_{A}^{(3)}  & =4\chi_{A}^{(3)}+\sqrt{2}[(s_{1x}\chi_{0}%
(s_{1z}))(s_{2x}\chi_{-1}(s_{2z}))-(s_{2x}\chi_{0}(s_{2z}))(s_{1x}\chi
_{-1}(s_{1z}))]\nonumber \\
& +\sqrt{2}[(s_{1y}\chi_{0}(s_{1z}))(s_{2y}\chi_{-1}(s_{2z}))-(s_{2y}\chi
_{0}(s_{2z}))(s_{1y}\chi_{-1}(s_{1z}))]\nonumber \\
& +\sqrt{2}[(s_{1z}\chi_{0}(s_{1z}))(s_{2z}\chi_{-1}(s_{2z}))-(s_{2z}\chi
_{0}(s_{2z}))(s_{1z}\chi_{-1}(s_{1z}))]\nonumber \\
& =4\chi_{A}^{(3)}+\sqrt{2}[\frac{1}{\sqrt{2}}(\chi_{1}(s_{1z})+\chi
_{-1}(s_{1z}))\frac{1}{\sqrt{2}}\chi_{0}(s_{2z})-\frac{1}{\sqrt{2}}(\chi
_{1}(s_{2z})+\chi_{-1}(s_{2z}))\frac{1}{\sqrt{2}}\chi_{0}(s_{1z})]\nonumber \\
& +\sqrt{2}[\frac{-i}{\sqrt{2}}(\chi_{1}(s_{1z})-\chi_{-1}(s_{1z}))\frac
{-i}{\sqrt{2}}\chi_{0}(s_{2z})-\frac{-i}{\sqrt{2}}(\chi_{1}(s_{2z})-\chi
_{-1}(s_{2z}))\frac{-i}{\sqrt{2}}\chi_{0}(s_{1z})]\nonumber \\
& =4\chi_{A}^{(3)}-2\chi_{A}^{(3)}\nonumber \\
& =2\chi_{A}^{(3)},\label{81}%
\end{align}
i.e.,
\begin{equation}
S=1,\label{82}%
\end{equation}
we have checked $\chi_{A}^{(3)}$ is the spin wave function of two-photon
corresponding to $S=1$ and $\mu=-1$.\newline By calculation and checking, we
obtain the correct spin wave function of two-photon, they are

(1) $S=2$ spin wave functions
\begin{equation}
\chi_{S}^{(1)}=\chi_{22}(s_{1z},s_{2z})=\chi_{1}(s_{1z})\chi_{1}%
(s_{2z}),\hspace{0.3in}(S=2,\mu=2)\label{83}%
\end{equation}%
\begin{equation}
\chi_{S}^{(2)}=\chi_{21}(s_{1z},s_{2z})=\frac{1}{\sqrt{2}}[\chi_{0}%
(s_{1z})\chi_{1}(s_{2z})+\chi_{0}(s_{2z})\chi_{1}(s_{1z})],\hspace
{0.3in}(S=2,\mu=1)\label{84}%
\end{equation}%
\begin{equation}
\chi_{S}^{(3)}=\frac{2}{\sqrt{6}}\chi_{0}(s_{1z})\chi_{0}(s_{2z})+\frac
{1}{\sqrt{6}}[\chi_{1}(s_{1z})\chi_{-1}(s_{2z})+\chi_{1}(s_{2z})\chi
_{-1}(s_{1z})],\hspace{0.3in}(S=2,\mu=0)\label{85}%
\end{equation}%
\begin{equation}
\chi_{S}^{(4)}=\chi_{2-1}(s_{1z},s_{2z})=\frac{1}{\sqrt{2}}[\chi_{0}%
(s_{1z})\chi_{-1}(s_{2z})+\chi_{0}(s_{2z})\chi_{-1}(s_{1z})],\hspace
{0.3in}(S=2,\mu=-1)\label{86}%
\end{equation}%
\begin{equation}
\chi_{S}^{(5)}=\chi_{2-2}(s_{1z},s_{2z})=\chi_{-1}(s_{1z})\chi_{-1}%
(s_{2z}),\hspace{0.3in}(S=2,\mu=-2)\label{87}%
\end{equation}
(2) $S=0$ spin wave function
\begin{equation}
\chi_{S}^{(6)}=\frac{1}{\sqrt{3}}\chi_{0}(s_{1z})\chi_{0}(s_{2z})-\frac
{1}{\sqrt{3}}[\chi_{1}(s_{1z})\chi_{-1}(s_{2z})+\chi_{1}(s_{2z})\chi
_{-1}(s_{1z})],\hspace{0.3in}(S=0,\mu=0)\label{88}%
\end{equation}
(3) $S=1$ spin wave functions
\begin{equation}
\chi_{A}^{(1)}=\chi_{11}(s_{1z},s_{2z})=\frac{1}{\sqrt{2}}[\chi_{0}%
(s_{1z})\chi_{1}(s_{2z})-\chi_{0}(s_{2z})\chi_{1}(s_{1z})],\hspace
{0.3in}(S=1,\mu=1)\label{89}%
\end{equation}%
\begin{equation}
\chi_{A}^{(2)}=\chi_{10}(s_{1z},s_{2z})=\frac{1}{\sqrt{2}}[\chi_{1}%
(s_{1z})\chi_{-1}(s_{2z})-\chi_{1}(s_{2z})\chi_{-1}(s_{1z})],\hspace
{0.3in}(S=1,\mu=0)\label{90}%
\end{equation}%
\begin{equation}
\chi_{A}^{(3)}=\chi_{1-1}(s_{1z},s_{2z})=\frac{1}{\sqrt{2}}[\chi_{0}%
(s_{1z})\chi_{-1}(s_{2z})-\chi_{0}(s_{2z})\chi_{-1}(s_{1z})],\hspace
{0.3in}(S=1,\mu=-1)\label{91}%
\end{equation}
From equations (\ref{83})-(\ref{91}), we can find the two-photon spin states
of $S=2$ and $S=0$ are symmetrical spin states, and the spin states of $S=1$
are antisymmetrical spin states. In the two-photon spin states, the states
$\chi_{S}^{(2)}$, $\chi_{S}^{(3)}$, $\chi_{S}^{(4)}$, $\chi_{S}^{(6)}$,
$\chi_{A}^{(1)}$, $\chi_{A}^{(2)}$ and $\chi_{A}^{(3)}$ are two-photon
entanglement states. We know the two-electron entanglement states are
\begin{equation}
\chi_{S}=\frac{1}{\sqrt{2}}[\chi_{\frac{1}{2}}(s_{1z})\chi_{-\frac{1}{2}%
}(s_{2z})+\chi_{\frac{1}{2}}(s_{2z})\chi_{-\frac{1}{2}}(s_{1z})],\hspace
{0.3in}(S=1,\mu=0)\label{92}%
\end{equation}
and
\begin{equation}
\chi_{A}=\frac{1}{\sqrt{2}}[\chi_{\frac{1}{2}}(s_{1z})\chi_{-\frac{1}{2}%
}(s_{2z})-\chi_{\frac{1}{2}}(s_{2z})\chi_{-\frac{1}{2}}(s_{1z})],\hspace
{0.3in}(S=0,\mu=0)\label{93}%
\end{equation}
where $\chi_{\frac{1}{2}}(s_{z})$ and $\chi_{-\frac{1}{2}}(s_{z})$ are $s_{z}$
eigenstates of eiegnvalues $\frac{1}{2}$ and $-\frac{1}{2}$, i.e., the
two-electron entanglement states are expressed by the spin state of two single
electron.\newline At present, the two-photon entanglement states are expressed
as \cite{s23,s24,s25,s26}
\begin{equation}
\chi_{1}=\frac{1}{\sqrt{2}}(|H\rangle_{1}|H\rangle_{2}\pm|V\rangle
_{1}|V\rangle_{2}),\label{94}%
\end{equation}
and
\begin{equation}
\chi_{2}=\frac{1}{\sqrt{2}}(|H\rangle_{1}|V\rangle_{2}\pm|V\rangle
_{1}|H\rangle_{2}).\label{95}%
\end{equation}
Where $|H\rangle$ and $|V\rangle$ denote the states of horizontal and vertical
linear polarization. We think equations (\ref{94}) and (\ref{95}) are
classical polarization vector expression, and are not complete. The quantum
expression of two-photon spin states should be expressed by the spin state of
photon, and the complete entanglement states of two-photon are shown in
equations (\ref{84}), (\ref{85}), (\ref{86}), (\ref{88}), (\ref{89}),
(\ref{90}) and (\ref{91}).

\section{Conclusion }

In a number of literature, the two-photon entanglement states are expressed by
equations (\ref{94}) and (\ref{95}), they are a classical polarization vector
expression and they are not complete. In this paper, we have given the spin
states of two-photon, which are expressed by the quadratic combination of two
single photon spin states. Obviously, they are a kind of quantum expression.
Otherwise, we give all two-photon entanglement states.

\end{document}